\documentclass[aps,prc,preprint,superscriptaddress]{revtex4}

\usepackage{graphics}

\begin{document}

\title{Spin-dependent observables in surrogate reactions}

\author{Satoshi CHIBA}
\email{chiba.satoshi@jaea.go.jp}
\affiliation{Japan Atomic Energy Agency, Tokai, Naka, Ibaraki 319-1195, Japan}

\author{Osamu IWAMOTO}
\affiliation{Japan Atomic Energy Agency, Tokai, Naka, Ibaraki 319-1195, Japan}

\author{Yoshihiro ARITOMO}
\affiliation{Japan Atomic Energy Agency, Tokai, Naka, Ibaraki 319-1195, Japan}

\date{\today}

\begin{abstract}

Observables emitted from various spin states in
compound U nuclei are investigated to validate usefulness of the
surrogate reaction method.  It was found that energy
spectrum of cascading $\gamma$-rays and their multiplicities, spectrum
of evaporated neutrons, and mass-distribution of fission fragments show
clear dependence on the spin of decaying nuclei.  The present results
indicate that they can be used to infer populated spin distributions
which significantly affect the decay branching ratio of the compound system
produced by the surrogate reactions.
\end{abstract}

\pacs{
24.87.+y, 
24.10.-i, 
24.60.Dr, 
25.85.Ec  
}
\keywords{neutron cross sections, surrogate
method, spin-dependence, observables, Hauser-Feshbach theory, Langevin
equation}

\maketitle

\section{Introduction}

Surrogate reaction is a method to measure neutron cross sections of
unstable or rare nuclei, for which preparation of enough amount of samples is
extremely difficult or practically impossible.
For example, neutron cross section of minor actinides,
long-lived fission products or branch-point nuclei of the s-process
nucleosynthesis may be measured without having them directly
but instead by preparing samples consisting of nascent stable isotopes.
Because of this nature, the surrogate method can be a unique tool to
investigate neutron cross sections of nuclei for direct measurement using
neutrons is difficult.  See, e.g., Ref.~\cite{kessedjian} for recent results
and references
therein.

By the surrogate method, the same compound nuclei
populated in desired neutron-induced reactions are generated by
(multi) nucleon transfer reactions, and decay branching ratios from them
are measured.  From this information, we can determine desired neutron
cross sections either 1) by directly multiplying total reaction cross section
to the branching ratio (surrogate absolute method) or 2) as a ratio
to a decay branching ratio of a compound system for which
neutron-induced cross section leading to it is known (surrogate ratio method,
 SRM).
Generally speaking, however, decay branching ratios are sensitive to the
spin value of the decaying compound system.  Therefore, a great care must
be paid to the condition in which the surrogate method can yield the
correct neutron cross sections\cite{scielzo,goldblum}.

Recently, SC and OI have shown that SRM
works if 1) spin distributions in 2 compound nuclei populated in the surrogate
ratio method are equivalent, 2) spin values populated by the surrogate
reaction is not much larger than
10 (in unit of $\hbar$), which is hardly populated in neutron-induced
reactions, and 3) J$^\pi$ by J$^\pi$ convergence of branching ratio
, called weak Weiskopf-Ewing condition,
is achieved\cite{chiba2010}.
The condition 1) is expected to be satisfied by selecting a pair of
nuclei having similar discrete level structure.
The condition 3) can be verified by a statistical model
calculation.
Then, the
condition 2) must be verified which is the
subject of this work.  These theoretical conditions, altogether,
then should be considered to design experimental equipments and
setups.

In this work, we have investigated 3 observables which are possibly
sensitive to the spin values of decaying nuclei, namely, energy
spectrum of evaporated neutrons, energy spectra and multiplicity of
emitted $\gamma$-rays and fission fragment mass distributions (FFMDs).
In the next section, computational methods are explained, which is followed
by their results and discussion.

\section{Computational Methods}

The details
of this calculation is given elsewhere\cite{CCONE,aritomo2011}, so only a
simple explanation is given here.

As explained in the introduction, it is enough to estimate
spin values in the step of approximately 10 $\hbar$.
We then took an example of
compound nucleus of $^{238}$U at excitation energy of 10 MeV, and
calculated spectra of emitted neutrons and $\gamma$-rays at $J^\pi$=0$^+$,
5$^+$, 10$^+$ and 20$^+$ by the Hauser-Feshbach theory\cite{hauser}.
Nonetheless to say, it corresponds to
neutron-induced reactions on $^{237}$U, or surrogate reaction such as
$^{236}$U($^{18}$O,$^{16}$O)$^{238}$U.  The calculation was
carried out by CCONE code system\cite{CCONE}.  Parameters used to
generate JENDL actinide file 2008\cite{JENDLact} were used.

Another calculation was also carried out.  The FFMD from $^{240}$U
was calculated by a 3-dimensional Langevin method on a potential
energy surface calculated by the two-center shell model\cite{maru72,sato78,aritomo2011,arit09}.
The total potential energy consists of a macroscopic rotating liquid-drop model which gives the bulk potential energy
for arbitrary spin states and a microscopic shell- and pairing-correction part:
\begin{equation}
V(q,J,T) = V_{LD}(q) + \frac{\hbar^2}{2 {\cal I}(q)} J(J+1) + V_{SH}(q,T)
\end{equation}
where $V(q,J,T)$ denotes the total potential energy
for deformation $q$, spin $J$ and temperature $T$,
$V_{LD}(q)$ potential energy of the finite-range liquid-drop model.
The symbol ${\cal I}(q)$ denotes the moment of
inertia of a rigid body at deformation $q$,
while $V_{SH}$ designates the shell- and pairing-correction
energy at temperature $T$.
The temperature is related to the excitation energy $E^*$ as
\begin{equation}
T = \sqrt{\frac{E^*}{a}}
\end{equation}
where $a$ denotes the level density parameter.  The shell effect has
a temperature dependence expressed by a dumping factor
\begin{equation}
\phi = e^{-\frac{E^*}{E_d}}.
\end{equation}
The shell dumping energy $E_d$ was chosen as 20 MeV \cite{igna75}, which was used in our previous
calculations\cite{aritomo2011}.
The deformation $q$
represents a set of parameters such as elongation, mass asymmetry and
fragment deformation used to express complicated shapes
which appear during the fission process.
The $J(J+1)$ dependence in the above formula
changes the liquid-drop energy, which
changes the smooth landscape of the potential energy surface on
which collective variables $q$ are driven as a function of time.
On the contrary, the shell correction does not depend on $J$
in our calculation.
The nucleus $^{240}$U was selected since we have a preliminary
experimental FFMD data, which was used to verify the general accuracy
of our calculation.

\section{Results and Discussion}

\begin{figure}
\begin{center}
\includegraphics{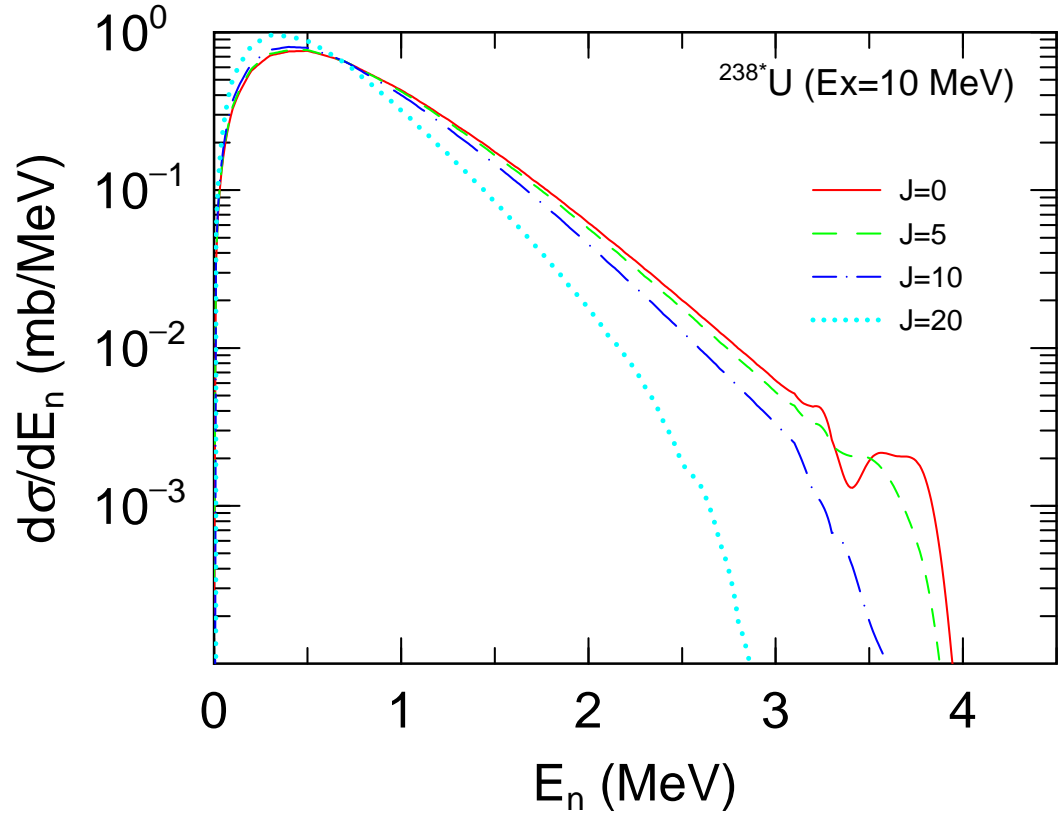}
\caption{(Color online) Energy spectra of evaporated neutrons from various spin states of
$^{238}$U at excitation energy of 10 MeV.}
\label{figure1}
\end{center}
\end{figure}

\begin{figure}
\begin{center}
\includegraphics{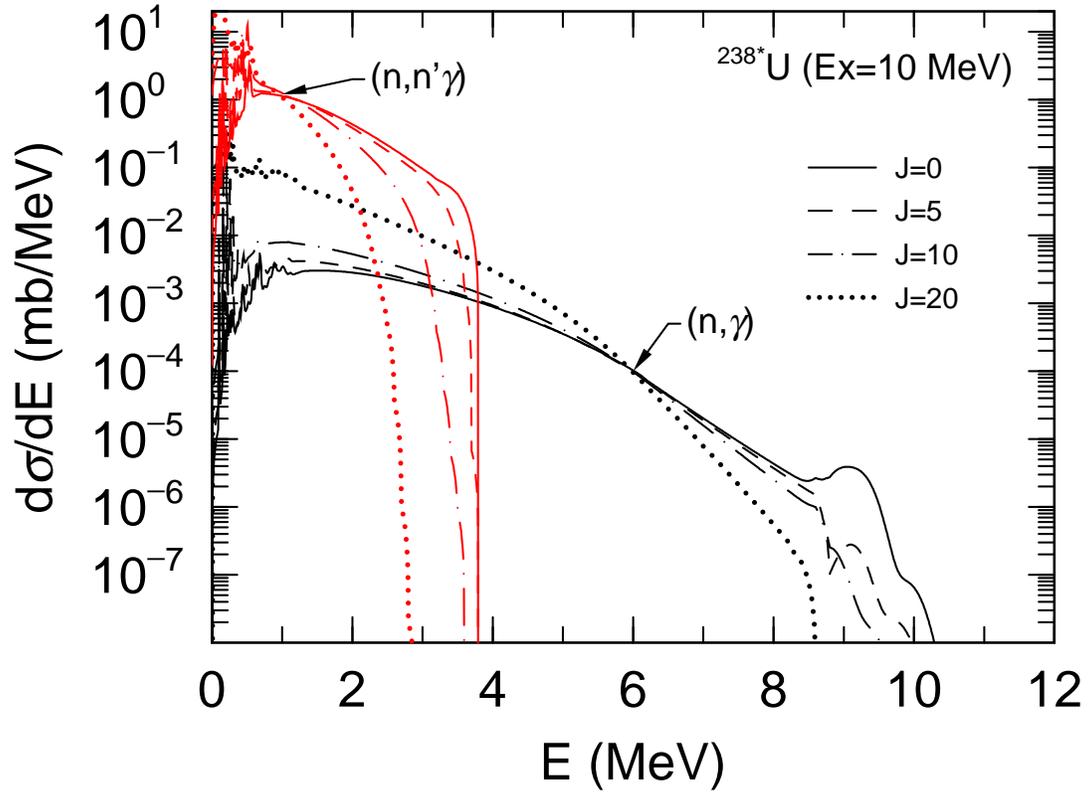}
\caption{(Coloe online)
Energy spectra of cascading $\gamma$-rays from various spin states of
$^{238}$U at excitation energy of 10 MeV.}
\label{figure2}
\end{center}
\end{figure}

The calculated energy spectra of evaporated neutrons and cascading
$\gamma$-rays are shown in Figs.~\ref{figure1} and \ref{figure2}, respectively.  They do not
include contributions from fission events, which can be eliminated
by applying veto in actual experiments.

In Fig.~\ref{figure1}, we notice that neutrons are emitted up to energy of about
4 MeV for low spin values.  The highest energy neutrons show some
irregular energy distributions reflecting the discrete level structure of
 $^{237}$U.  When the spin is increased to 10, the maximum energy
is decreased to slightly less than 3.8 MeV, and the spectra becomes
smoother.  For $J$=20, the maximum neutron energy is further reduced
to 3 MeV.  The decrease of the maximum neutron energy comes from the
fact that the level densities having large spins are rather low below
excitation energy around 4 MeV.  Therefore, if the surrogate reactions
produces only the compound states having spin more than 10, it will be
signaled by energy spectra of evaporated neutrons softer than
those of normal neutron-induced reactions (or corresponding statistical
model calculation).

\begin{figure}
\begin{center}
\includegraphics{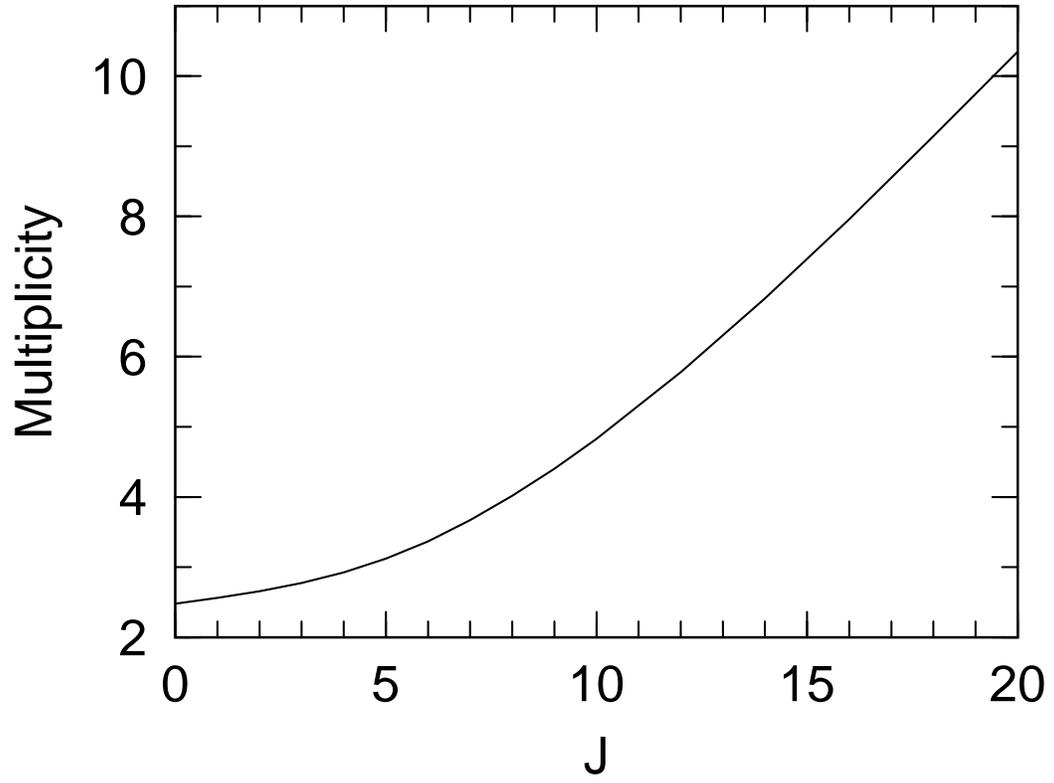}
\caption{Multiplicities of $\gamma$-rays emitted from
various spin states of
$^{238}$U at excitation energy of 10 MeV.}
\label{figure3}
\end{center}
\end{figure}

The same trend is observed for the energy spectra of cascading $\gamma$-rays.
In Fig.~\ref{figure2}, both the $\gamma$-rays
leading to ground-states of $^{238}$U and $^{237}$U are shown separately as
(n,$\gamma$) and (n,n'$\gamma$) spectra.  Difference of the $\gamma$-ray
energy spectra for both reactions at $J$=0 and 10 are visible.
The spin-dependence clearly affects the total $\gamma$-ray multiplicities,
which is shown in Fig.~\ref{figure3}.  $J$=0 gives multiplicity of about 2, while
it increases to 5 for $J$=10, and more than 10 for $J$=20.  This is
due to the fact that higher-energy $\gamma$
transition leading to low-lying levels is hindered when $J$ is large
due to the fact that high $J$ states are rare at low excitation energy.

\begin{figure}
\begin{center}
\includegraphics{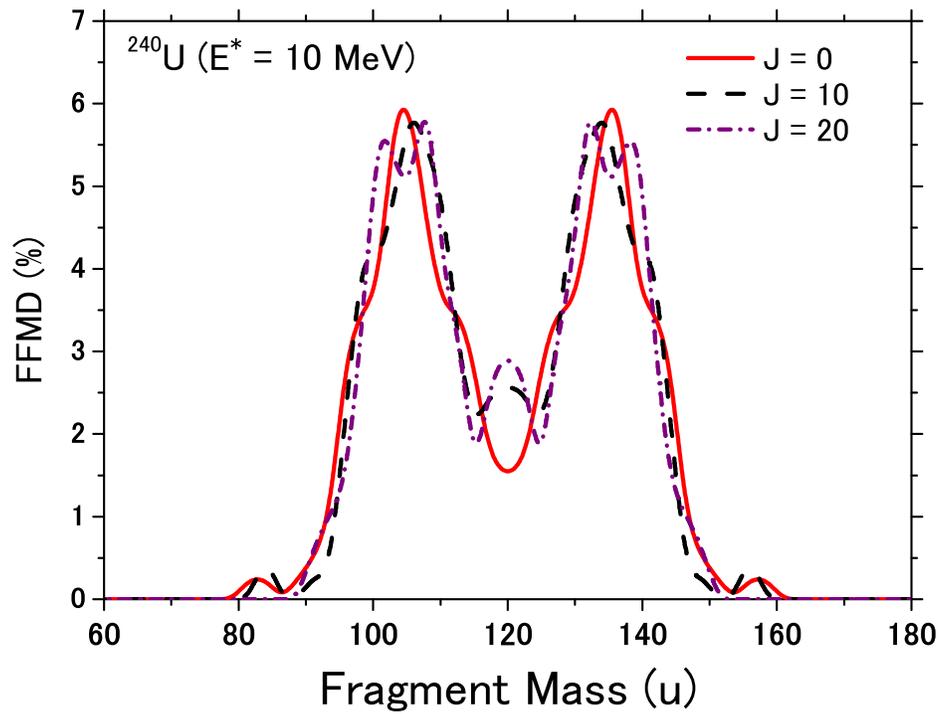}
\caption{(Color online) Fission fragment mass distribution (FFMD) emitted from
$J$=0, 10 and 20 states of $^{240*}$U
at excitation energy of 10 MeV.}
\label{figure4}
\end{center}
\end{figure}

The FFMD from $^{240*}$U is shown in Fig.~\ref{figure4}.
In this mass region, FFMD
show typical asymmetric distribution for low $J$ values,
which is well reproduced by our
calculation.  Then, we put $J$=10 and 20, and generated many Langevin
trajectories and compared them with the $J$=0 results.
We can notice that the peak of the
asymmetric mass division does not change, but the symmetric components
are sightly enhanced when $J$ becomes 10 and 20.  It can be interpreted that
when $J$ becomes larger, the liquid-drop energy, which gives minimum
at the symmetric division, increases, and the overall effect is to
enhance the symmetric fission.
If this is really the
case, the FFMD data, which can be obtained in our forthcoming experiments,
can be used to estimate the spin distributions populated in surrogate reactions.
However, the spin-dependence of the FFMD pointed out here is not a
phenomena confirmed yet.  Therefore we must place a special care on
this result, although it may give a new insight into the understanding
of the fission mechanism.

\section{Concluding remarks}

We have investigated possible observables which can signal the
spin (distribution) of compound nuclei populated in surrogate reactions.
The energy spectra of evaporation neutrons and cascading $\gamma$-rays
clearly show difference for spin states of $J$=0 and 10 or above.
Generally, the high-spin states produces softer neutron and $\gamma$-ray
spectra.  This is due to the spin-dependence of level density.  The
$\gamma$-ray multiplicity increases as the spin of decaying compound
nucleus increases since high-energy transitions leading to low-spin
states near the ground state are hindered.  We also found that fission
fragment mass distribution may have spin dependence.  This last point
may become an interesting issue from the viewpoint of fission physics.
We understand that it may not be easy to measure evaporation neutrons
and cascading $\gamma$-rays in the presence of fission events.  If so,
we can use lighter, non-fissioning target to measure these quantities.
These results, altogether, are used to
design the forthcoming experiments and understanding the surrogate
reaction and its relation to desired neutron-induced reaction will be
enhanced.

\begin{acknowledgments}
The authors are grateful to Dr. K. Nishio for helpful discussions.
Present study is the result of
``Development of a Novel Technique for Measurement
of Nuclear Data Influencing the Design of Advanced Fast Reactors"
entrusted to Japan Atomic
Energy Agency (JAEA) by the
Ministry of Education, Culture, Sports, Science and Technology of Japan
(MEXT).
\end{acknowledgments}


\begin{thebibliography}{99}


\bibitem{kessedjian}G.~Kessedjian {\it et al.}, Phys. Lett.
\textbf{B 692}, 297-301(2010).

\bibitem{scielzo}N.D.~Scielzo {\it et al.}, {Phys. Rev. C}
{\bf 81}, 034608-1-12(2010).

\bibitem{goldblum}
B.L.~Goldblum, S.G.~Prussin, L.A.~Bernstein, W.~Younes, M.~Guttormsen, and H.T.~Nyhus, {Phys. Rev. C} {\bf 81} 054606-1-7(2010).

\bibitem{chiba2010} S.~Chiba and O.~Iwamoto, {Phys. Rev. C}, \textbf{81}, 04604-1-6 (2010).

\bibitem{CCONE} O.~Iwamoto, {J. Nucl. Sci. Technol.}, \textbf{44},
678(2007).

\bibitem{aritomo2011} Y.~Aritomo, S.~Chiba and K.~Nishio,
arXiv:1009.5924v1 (2010) [nucl-th].

\bibitem {hauser}W.~Hauser and H.~Feshbach,
{Phys. Rev.} \textbf{87}, 366-373(1952).

\bibitem{JENDLact} O.~Iwamoto, T.~Nakagawa, N.~Otsuka, S.~Chiba,
K.~Okumura, G.~Chiba, T.~Ohsawa and K.~Furutaka, {J. Nucl. Sci.
Technol.}, \textbf{46}, 510(2008).

\bibitem{maru72}J.~Maruhn and W.~Greiner, Z. Phys. A {\bf 251}, 431 (1972).
\bibitem{sato78}K.~Sato, A.~Iwamoto, K.~Harada, S.~Yamaji, and S.~Yoshida,
 Z. Phys. A {\bf 288}, 383 (1978).


\bibitem{arit09} Y.~Aritomo, Phys. Rev. C {\bf 80} (2009) 064604.
\bibitem{igna75} A.N.~Ignatyuk, G.N.~Smirenkin, and A.S.~Tishin, Sov.
J. Nucl. Phys. {\bf 21}, 255 (1975).



\end{thebibliography}

\newpage

\end{document}